\magnification=1100
\centerline{\bf Three-Dimensional Modeling of the Deflagration Stage}
\centerline{\bf of a Type Ia Supernova Explosion}

\bigskip
\centerline{A. M. Khokhlov}

\bigskip 
\centerline{Laboratory for Computational Physics and Fluid Dynamics,}
\centerline{Naval Research Laboratory, Washington, DC 20375}

\bigskip\centerline{\bf Abstract}

The paper describes a physical model and  numerical algorithm for modeling Type Ia supernova (SNIa)
explosions in three-dimensions and presents first
results of modeling a deflagration explosion in a non-rotating, Chandrasekhar-mass carbon-oxygen (CO)
white dwarf. Simulations show that the turbulent flame speed grows exponentially, 
reaches  $\simeq30$\% of the  speed of sound, and then declines as the  
large-scale turbulence is frozen  by expansion. The freezing of turbulent motions appears to be a
crucial physical mechanism regulating the rate of deflagration in SNIa. The
energy of the explosion is comparable to that of a typical SNIa. However, the presence of the
outer layer of unburned CO and the formation of intermediate mass elements and pockets of unburned CO near
the center  pose problems for the  modeling of SNIa spectra. Delayed detonation is a way to
alleviate these problems and to produce consistent spectra.

\vfill\eject

\bigskip\noindent
\bigskip\centerline{\bf 1. Introduction}

\bigskip
Type Ia Supernovae (SNIa) are thought to be thermonuclear explosions of
Chandra\-sekhar-mass carbon-oxygen white dwarfs (CO-WD). They play an important
role in nucleosynthesis (Nomoto et al. 1984) and the balance of energy in interstellar medium, and they
are important distance indicators used in cosmology (Perlmuter et al., 1999; Riess et al. 2000). Our
understanding of SNIa explosions is far from complete. To predict the explosion  outcome, 
one needs to know
how  thermonuclear burning propagates inside the exploding star. The mechanisms and the
speed of thermonuclear burning in SNIa remain an unsolved theoretical problem. A recent review
by  Hillebrandt \& Niemeyer (2000) contains a large list of relevant publications. 
 
Burning can propagate in a supernova either as a flame caused by the electron
heat conduction or as a detonation in which reactions  are triggered by a shock.
Both the detonation speed and the laminar flame  speed $S_l$ in SNIa are
known.  However, in the
gravitational field of the star, the laminar  flame  becomes  unstable with respect to the
Rayleigh-Taylor (RT) instability (Nomoto et al. 1976). This  leads to
 a turbulent deflagration with some unknown turbulent flame speed $S_t > S_l$. 

Observations indicate that burning of a Chandra\-sekhar-mass  CO-WD must start as a deflagration. Spectra
of SNIa at maximum light show the presence of large amounts of intermediate mass elements such as Si, S,
Mg (Pskovskii 1977, Branch 1981). These elements
 can
only be synthesized at densities less than $\simeq 10^7$g/cc, where the temperature of
burned products stays below $\simeq (4-5) \times 10^9$K.  When burning  starts  as a
deflagration, the initially high-density WD expands while it
burns,  its density  decreases, and this  creates the necessary
conditions for the production of intermediate mass elements. 
In one-dimensional deflagration models, $S_t$ is a free parameter (Nomoto et al. 1976, Woosley \& Weaver
1986). Delayed detonation models assume  that a deflagration undergoes a transitions to a detonation at
low densities (Khokhlov 1991). In these models, both $S_t$ and  the moment of a deflagration-to-detonation
transition (DDT) are  free parameters.  To fit observations,  burning  must be initially slow in order to 
expand the  WD significantly. Then burning must  accelerate quickly to incinerate    expanding
outer layers rapidly. 
 This requires  very carefully tuning  $S_t$ as a function of time in deflagration
models. Delayed-detonation models accomplish the incineration  by  
 switching to a supersonic  detonation mode of burning. 

Two-dimensional simulations of SNIa explosions 
show that burning leads to a significant expansion of a WD. However, the deflagration speed found in all
of the simulations is  too small to lead to a powerful  explosion 
(Livne, 1993;  Arnett \& Livne, 1994; Reinecke et al., 2000). Whether this is an inherent difficulty
of the deflagration model or simply a deficiency of two-dimensional simulations is a question.
Two-dimensional simulations lack a key physical ingredient of turbulent burning -- turbulent energy
cascade that  is directed in three dimensions from  large to small spatial scales. Simulations of burning
in a vertical column in a uniform gravitational field show that the turbulent flame speed grows faster in
three dimensions and eventually reaches a turbulent steady state where
$S_t$ becomes independent of $S_l$ and of the details of burning on small scales 
(Khokhlov, 1995; hereafter
K95). In two dimensions, $S_t$ grows more slowly and depends on the value of $S_l$. However, turbulence in
the exploding star is affected by expansion on  spatial scales
$\cal L$, where the characteristic RT timescale 
$\tau_{RT}\simeq (g{\cal L})^{1/2}$ is comparable or exceeds the expansion timescale
 $\tau_e\simeq R/U_e$, where $R$ is stellar radius and $U_e$ is the
(time-dependent) expansion velocity. Expansion  tends to freeze large-scale
turbulent motions and thus decreases the turbulent flame speed (K95). 
The actual $S_t$ in a supernova must depend on  the
competition of  the RT instability, turbulent energy cascade, and 
expansion. To predict $S_t$, these effects must be modeled in three dimensions.
This paper describes a physical model and  numerical algorithm for modeling SNIa
explosions in three-dimensions  and presents first
results of modeling a deflagration explosion in a nonrotating, Chandrasekhar-mass CO-WD.

\bigskip\noindent
\bigskip\centerline{\bf 2. Physical and numerical model}

\bigskip\centerline{\it 2.1. Fluid dynamics}

\bigskip
Stellar matter is described by the  Euler equation for an inviscid
fluid
$$
\eqalign{
{{\partial\rho}\over{\partial t}} &= - \nabla\cdot\left(\rho {\bf U}\right)~,\cr
{{\partial\rho{\bf U}}\over{\partial t}} &= - \nabla\cdot\left(\rho {\bf U}{\bf U}\right)
             -\nabla P + \rho {\bf g} ~,\cr
{{\partial E}\over{\partial t}} &= - \nabla\cdot\left({\bf U}\left( E+P\right)\right) +
    \rho {\bf U}\cdot{\bf g} + \rho \dot q~, \cr 
}   \eqno(1)
$$
where $\rho$, $E = E_i + \rho U^2/2$, $E_i$,  ${\bf U}$, $\bf g$ and $\dot q$ are the mass density, energy
density, internal energy density,  velocity of matter,  gravitational acceleration, and nuclear  energy
release rate per unit mass, respectively. The acceleration of gravity is computed using an angle-averaged
density distribution. 
The equation of state (EOS) of degenerate matter includes contributions from
ideal Fermi-Dirac electrons and positrons, equilibrium Planck radiation, and ideal ions.
Pressure $P = P(\rho,E_i,Y_e,Y_i)$ and temperature
$T=T(\rho,E_i,Y_e,Y_i)$ are determined by the EOS as functions of
$\rho$, $E_i$, electron mole fraction
$Y_e$, and the mean mole fraction of ions $Y_i$.   

\bigskip\centerline{\it 2.2. Flame propagation}

\bigskip
The flame is advanced using a flame-capturing technique which mimics
a flame with a prescribed normal  speed (K95). 
 A scalar variable $f$, such that  
$f=0$ in the unburned matter and $f=1$ in the material which has passed through the
flame, obeys a reaction-diffusion equation 
$$
 {{\partial f}\over{\partial t}} + {\bf U}\cdot\nabla f = K \nabla^2
f + R~, 
\eqno(2)
$$
with artificial reaction and diffusion coefficients 
$$
\eqalign{
R&= \cases{ C={\rm const.}, &if $f_0\leq f \leq 1$;\cr
            0, &otherwise,\cr}\cr
K&={\rm const}, \cr
}    \eqno(3)
$$
with some $0<f_0<1$.   In this paper, 
$f_0=0.3$. Equation (2) has a solution $f(x-St)$ which 
 describes a reaction front that
propagates with the  speed 
$S= (KC/f_0)^{1/2}$ and  has a thickness
$\delta
\simeq (K/C)^{1/2}$ (see  Appendix of (K95)).
If $K$ and $C$ 
in (3) are set to 
$$
K = S  (\beta \Delta r) \sqrt{f_0} ~,~~~~~
C = {S \over ( \beta \Delta r) } \sqrt{f_0}~, \eqno(4)
$$
with $\beta={\rm constant}$, the front    
propagates  with a
prescribed speed
$S$ and  spreads onto several  computational cells of size $\Delta r$.  
The choice of  $\beta=1.5$ spreads the flame  on $\simeq  3-4$
cells. Making the front
 narrower  is not practical, since the fluid dynamics algorithm spreads
contact discontinuities on $\simeq 4$ cells. 
The energy-release rate inside the front is defined as 
$$
\dot q =  \,q_f {df\over dt}~, \eqno(5)
$$
where  $q_f$, defined below in Section 3.4, is the  nuclear energy release inside the front.
Due to the energy release, both density
and velocity vary across the front, and fluid motions are generated which advect the flame relative to
the computational mesh. The 
flame-capturing technique described above advances the reaction front relative to the fuel
with the speed practically independent of the orientation of the front on the mesh, fluid motions, and 
resolution. 

\bigskip\centerline{2.3. Burning on small scales}

\bigskip
A turbulent flame in a vertical column of width $\cal L$ subjected to an acceleration of
gravity
$g$ propagates in a steady-state with the speed
$
S_t \simeq 0.5 \sqrt{ Ag{\cal L}}        
$
independent of the laminar speed $S_l$, where $A=(\rho_0 - \rho_1)/( \rho_0 + \rho_1)$ is
the Atwood number, and $\rho_0$, $\rho_1$ are  the densities ahead and behind the flame
front, respectively (K95). The time to reach a steady state is  $\simeq 2{\cal L}/S_t$,
and the thickness of the flame brush is  $\simeq 2 \cal L$. 
To describe  burning on scales that are not resolved in the simulations, the value of
$S$ in (4) is taken as
$$
S= \max ( S_l, S_{sub} )~,  \eqno(6)
$$
with  
$$
S_{sub} = 0.5 \sqrt{  A\, \alpha \Delta r \, \left( s_1 \vert {\bf g}\vert
           + s_2 \max
             \left( 0, {\bf n}\cdot {\bf g} \right) \right)}~\eqno(7)
$$
based on the assumption that burning on small scales is driven by
the RT instability at these scales and that at ${\cal L} << R_{WD}$ burning adjusts to a  local
steady state (Livne 1993, Arnett
\& Livne 1994, K95).  In (7), $\alpha\simeq 1$ determines the driving scale, 
${\cal L} \simeq \alpha \Delta
r$, in terms of the computational cell size $\Delta r$, 
and ${\bf n} = \nabla f / f$ is the unit vector
normal to the flame front and directed towards the products of burning.
The choice of $s_1=1$,
$s_2=0$ gives a subgrid flame speed that is independent of the
orientation of the flame front relative to the direction of gravity. The choice
$s_1=0$, $s_2=1$ gives a directional-dependent speed that reduces to
$S_l$ when the flame moves in the direction of gravity. Finally, $s_1=s_2=0$ gives the
laminar flame speed.  In this paper, only the limiting,  angle-independent and laminar prescriptions for
$S$ were used. The Atwood number  $A \simeq   {(\gamma-1) \,{\bar \rho}\,q_f \,/ \,2 \gamma P}$ was
calculated from  conservation of enthalpy,  assuming constant pressure across the flame, 
where $\gamma$  and $\bar \rho = (\rho_1 + \rho_0)/2$ are average values of the
adiabatic index and mass density.   Under the  conditions of interest here, $A\simeq 0.1-0.4$.
The laminar flame speed
$S_l$,  taken according to Timmes \& Woosley (1992) and Khokhlov et al. (1997),  is 
$S_l \simeq (10^{-2} - 10^{-4})\times a_s$ for 
$\rho\simeq 10^9 - 10^7$g/cc,
where $a_s\simeq 5\times 10^8$cm/s is the sound speed.

\bigskip\centerline{\it 2.4. Nuclear kinetics}

\bigskip
There are three distinct stages of carbon
burning. First,  the  $^{12}C + \, ^{12}C$ reaction leads to the consumption  of
$^{12}C$ and formation of mostly $^{20}Ne$, $^{24}Mg$, protons and
$\alpha$-particles. Then the onset of the nuclear statistical quasi-equilibrium (NSQE)
takes place,  during which Si-group (intermediate mass) elements are formed.  Finally,
Si-group elements are converted into the Fe-group elements and the nuclear statistical
equilibrium (NSE) sets in.  The NSQE and NSE relaxation involves hundreds of species from
carbon to zinc and thousands of  reactions of these nuclei with  protons, neutrons,
and $\alpha$-particles.  The timescales of all these stages strongly depend on
temperature. The timescale of carbon consumption is much shorter than the
quasi-equilibrium timescale 
$\tau_{nsqe}$ which, in turn, is much shorter than the NSE timescale $\tau_{nse}$.

A four-equation kinetic scheme  is used here to describe
the energy release, synthesis of Si- and Fe-group elements, and the
neutronization of NSE matter. A similar scheme has been used in
one-dimensional  simulations of SNIa  (Khokhlov 1991).
The kinetic equation for the mole  fraction of carbon
$Y_C$
$$
{dY_C\over dt} = -\rho A(T_9) \exp(-Q/T_{9a}^{1/3}) Y_C^2 
~,\eqno(8)
$$
 describes carbon consumption  through the major reactions
$^{12}C \, (\,{^{12}C},\,p) \, ^{23}Na \,
(\,p,\gamma) \, ^{24}Mg$ and $^{12}C\,(\,{^{12}C},\,{^4He})\,^{20}Ne$  with the branching ratio
$\simeq 1$. Here
 $Q=84.165$, $T_{9a} = T_9/(1+0.067T_9)$, where $A(T_9)$ is a known function (Fowler et al.
1975).

Most of the nuclear energy is released during the
carbon exhaustion stage and  the subsequent synthesis of the 
Si-group nuclei. The energy release or consumption due to the transition from Si-group to
Fe-group nuclei (NSE relaxation) is  less than 10\%. Therefore, the nuclear energy release
rate is approximated as 
$$
{dq_n \over dt} = -Q_C {dY_C \over dt} +  { q_{nse} - q_n \over \tau_{nsqe} }
  ~,
\eqno(9)
$$
where $q_n(t)$ is the binding energy of nuclei per unit mass, $Q_C = 4.48\times
10^{18}$ ergs g$^{-1}$ mol$^{-1}$ is the energy release due to carbon burning (8),
and $q_{nse}(\rho,T,Y_e)$ is the binding energy of matter in the state of NSE.

The  equation 
$$
{d\delta_{nse}\over dt} = { 1 - \delta_{nse} \over \tau_{nse}}~,    \eqno(10)
$$ 
traces the onset of NSE, where $\delta_{nse} = 0$ in the unburned matter and 
$\delta_{nse} = 1$ in the NSE products. Intermediate mass elements  are expected
 where $Y_C(t=\infty) \simeq 0$ and $\delta_{nse}(t=\infty) < 1$.  Fe-group
elements are expected  where $Y_C(t=\infty) \simeq 0$ and
$\delta_{nse}(t=\infty) \simeq 1$.  
The `e-folding' NSQE and NSE timescales  
$$
\eqalign{
\tau_{nsqe} &= \exp \left( 149.7/T_9 -39.15 \right)~{\rm s} ~,\cr
\tau_{nse} &= \exp\left( 179.7/T_9 -40.5 \right)~{\rm s}~, \cr
}         \eqno(11)
$$
approximate the results of the detailed
calculations of carbon burning. The original  kinetic scheme (Khokhlov 1991) also
contained the equation for the mean ion mole fraction $Y_i$.  To simplify the scheme in this
paper,  variations of
$Y_i$ due to nuclear reactions were neglected  since ions make a 
small contribution to the equation of state.  A constant $Y_i=0.07$ was used instead,
which is an  average of $Y_i$ in the unburned and typical NSE matter in the SNIa
explosion conditions.

Neutronization is described by the equation for the
 electron mole fraction $Y_e$,
$$
{dY_e \over dt} = - R_w(\rho,T,Y_e)~.  \eqno(12)
$$
The corresponding term describing  neutrino energy losses, $-\dot q_w(\rho,T,Y_e)$, is
added to the energy conservation equation in (1). Values of $q_{nse}$,
$\dot q_w$, and $R_w$ were computed assuming  NSE distribution of individual nuclei, as
described in (Khokhlov 1991). Recently, there has been an important development in theoretical
computations  which shows significantly smaller  electron capture and $\beta$-decay
rates in stellar matter (Langanke et al. 1999   and references therein). Following Martinez-Pinedo et
al. (2000),  this effect was  approximately taken  into
account by decreasing
$\dot q_w$ and 
$R_w$ by a factor of $5$. This should  be sufficient to  account for changes in
$q_{nse}$ and in the corresponding  nuclear energy release $q_{nse} - q_n(t=0)$ caused by variations of
$Y_e$. In future simulations, we plan to use $\dot q_w$ and $R_w$  based on the
new individual capture rates.

Despite its simplicity, the kinetic scheme
reasonably well  describes all major stages of carbon burning. In particular, it 
takes into account  the important effect of energy release or absorption caused 
by changes in NSE composition  
 when matter expands or contracts (these changes happen on a quasi-equilibrium
rather than on the equilibrium timescale). When the WD expands, this effect  adds   
$\simeq 50$\% to the energy initially released by burning  at high densities. Using
even the simplest 13-species
$\alpha$-network to account for this effect  would be prohibitively expensive in
three-dimensional hydrodynamical simulations. 

 To couple the  kinetic scheme and the front tracking algorithm (Section 2.2), 
$q_f$ in (5) is taken 
$q_f = q_{nse} - q_n(0)$ at densities 
$\rho>2\times 10^7$ g/cc, $q_f = Q_C Y_C(t=0)$ at  $\rho<5\times 10^6$ g/cc, and
is linearly interpolated between the two values in the density range $5\times 10^6 -
2\times 10^7$ g/cc. The carbon mole fraction $Y_C$ inside the
front is decreased in proportion to the increase of $f$.  Thus, 
 at high densities, both carbon consumption and the NSQE relaxation take place inside
the flame front. At   low densities where the NSQE stage of burning is
slow,  NSQE relaxation    takes place outside the flame front.

\bigskip\centerline{\it 2.5. Numerical method}

\bigskip
Fluid dynamics equations (1) for $\rho$, {\bf U}, $E$, and  advection parts of
 equations (2), (8)-(10) and 912) for 
chemical variables  $f$, $Y_C$, $q_n$,
$\delta_{nse}$ and $Y_e$ are solved using an explicit, second-order, Godunov-type,
adaptive mesh refinement algorithm ALLA (Khokhlov, 1998). 
A Riemann solver for an arbitrary EOS  (Colella
\& Glaz, 1985) is  used to evaluate fluxes at cell interfaces. Left and right
states  are computed using a monotone  piecewise-linear reconstruction  (van Leer 1979). 
A small amount of an artificial
diffusion flux (Lapidus, 1967) is added to Euler fluxes following Colella and Woodward
(1984). This feature is used  only in the  vicinity of strong shocks. 
Multidimensions are treated using  directional splitting.  The diffusion term in (2) is treated
 using  an explicit, second-order finite differencing. 

ALLA uses a fully threaded tree (FTT) adaptive mesh refinement  to refine the
computational  mesh  on the  level of individual cells
(Khokhlov 1998; Khokhlov \&
Chtchelkanova, 1999). FTT is a tree with inverted
thread pointers directed from children to  neighbors of parent cells.
As opposed to conventional tree structures,  all operations on the FTT,
including tree modifications  (refinement and unrefinement) are
$O(N)$, where N is a number of cells. All of these operations can be performed in parallel.  FTT has a
$2{1\over2}$ integers-per-cell memory overhead,  compared with 8 integers for an ordinary
tree or 14 integers for a tree that stores connectivity information. Computations using the FTT are
virtually as fast (per computational cell) as  those on a regular grid. 
A massively parallel version of the code uses a space-filling curve approach to maintain data locality,
achieve load balance, and  minimize interprocessor communications. 
 ALLA has been extensively
tested and used in various combustion problems involving
shocks, flames, turbulence, and their interactions (Gamezo et al., 1999,2000;  Khokhlov \&
Oran, 1999; and references therein) and in astrophysics
(Khokhlov et al., 1999).

Mesh refinement was used   
to refine shocks, contact discontinuities, and  regions with large gradients of
$\rho$, $P$, and $f$. The expressions for the corresponding shock, $\xi_s$, discontinuity,
$\xi_c$, and gradient refinement indicators, 
$\xi_P$,
$\xi_\rho$, $\xi_f$, are given by the equations (11), (12) and (13) of Khokhlov (1998),
respectively. 
In addition, refinement  was introduced for the gradients of 
tangential velocity using an indicator
$$
\xi_u = \max_{j,k=1,...,3} \left(\, (1-\delta_{jk})\left(\partial
U_k\over\partial x_j
\right)     
\right) {\Delta x\over \alpha \,a_s} ~ \eqno(13)
$$
 with $\alpha=0.05$. Refinement was initiated in places where the maximum of all
 indicators,
$$
\xi = \max ( \xi_s, \xi_c, \xi_P, \xi_\rho, \xi_f, \xi_u )~, \eqno(14)
$$
was larger than  the threshold value $\xi > 0.5$. The mesh was unrefined in
places where $\xi$ was less than the threshold value $\xi < 0.05$. 
The cell size $\Delta r$ is related to the level $l$ of the cell in the tree as
$\Delta r= L/2^l$. The mesh can be characterized
by the minimum and maximum  levels of leaves (unsplit cells) in the tree,
$l_{min}$ and
$l_{max}$, which were  predefined at the beginning of a computation. During a computation, mesh refinement
was allowed for cells with $l_{min} \leq l < l_{max}$.

Computations were carried out on a 128 $250$MHz R10000  SGI Origin 2000 at the Naval Research
Laboratory (NRL). On this machine, ALLA advanced   $\simeq
2\times 10^4$ cells at the rate of one timestep per second  per processor  on  pure
fluid-dynamic, three-dimensional problems with $\gamma$ held constant. 
With the addition of
a degenerate matter EOS, flame capturing, and nuclear kinetics, the speed
decreased to $\simeq 6\times 10^3$ cell-steps/s/processor. 

\bigskip\centerline{\it 2.6. Problem setup and initial conditions}

\bigskip
Most of the   computation were
performed for one octant of the WD assuming mirror symmetry along the $x=0$, $y=0$ and $z=0$ planes
passing through the WD center. The initial temperature was taken as  $T=10^5$K everywhere inside the WD.
The initial central density was taken as $\rho_c=2\times 10^9$ g/cc. Starting from the central pressure
$P(\rho_c)$, the equations of hydrostatic equilibrium, $dP/dr = -GM\rho/r^2$ and
$dM/dr=4\pi\rho r^2$,  were integrated outward until
$P=0$ was reached. The resulting WD configuration was interpolated onto a three-dimensional mesh.
Nuclear kinetic variables for the unburned matter and
$f=0$ were defined everywhere.
To initiate burning in the center, $f=1$ was then
 set  
 inside a small sphere of radius $r_i = 3\times 10^6$ cm and mass
$M_i\simeq 8\times 10^{-5} M_{WD}$. For a while, burning was not allowed inside the sphere  in order not
 to impulsively disturb the hydrostatic equilibrium. This procedure smoothly  triggered the
flame propagation process just  outside the sphere.   Due to the finite numerical resolution, the shape of
the initial flame bubble was not perfectly spherical but contained a number of perturbations. A quadrupole
mode was the largest mode of perturbations consistent with the symmetries of the problem.
Smaller perturbations were mostly burned out by the flame, but the quadrupole mode partially survived 
and gave rise to the development of the RT instability.

\bigskip\centerline{\bf 3. Results}

\bigskip\centerline{\it 3.1. Hydrostatic equilibrium}

\bigskip
Before calculating flame propagation,  tests were made to
ensure  that without burning, the WD  remains hydrostatic and  keeps
its symmetry. Since a Chandrasekhar-mass WD is   close to a collapse 
threshold,  its equilibrium is very sensitive to  the discretization errors. Thus a rather fine mesh 
is required near
the WD center   to maintain  equilibrium.  
The initial mesh  was constructed with fine cells
at level
$l=l_{max}=10$  for radii $r < 0.4 R_{WD}$, at 
$l=l_{max}-1$ for $0.4 R_{WD} \le r < 0.8 R_{WD}$, at 
$l=l_{max}-2$ for $0.8 R_{WD} \le r < 1.1 R_{WD}$, and coarser cells at larger $r$.
Computations of   hydrostatic equilibrium with  $\Delta r(l_{max})=
5\times 10^5$ cm showed numerical noise 
$ < 3\times 10^5$cm/s everywhere except in the very outer layers of the WD where the density changed from
cell to cell by more than an order of magnitude. Near the center, the  noise was less than the fluid 
velocity generated by the laminar flame, 
$\simeq S_l (\rho_0-\rho_1)/\rho_0 \simeq 2\times 10^6$ cm/s.
The total kinetic energy $E_k$ of the WD
 remained
$E_k < 10^{-5} E_{th}$, where $E_{th}$ is the total internal energy,  during  
 more than two sound crossing times of the WD ($\simeq 0.9$ s of integration). Deviations from
spherical symmetry were negligible.

\bigskip\centerline{\it 3.2. Deflagration of a  $0.5C + 0.5O$ WD}

\bigskip
Deflagration   of a CO-WD with  initial  composition $X_C=0.5$ by mass ($Y_C = 1/24$),
central density $\rho_c=2\times 10^9$g/cc, and angle-independent small scale burning
$s_1=1$, $s_2=0$ and $\alpha = 1$ in  (7), serves as the baseline case. 
The computational domain for this simulation had  size $L=5.35\times 10^8$ cm  with levels of
refinement 
$l_{max}=10$,
$l_{min}=7$, and  with maximum and minimum resolution $\Delta r(l_{max})=5.22\times 10^5$  and  $\Delta
r(l_{min})=6.18\times 10^6$ cm, respectively. The simulation was carried out for  $\simeq 1.8$s of physical
time using   Courant number $0.7$, required $\simeq 9,000$ timesteps, and took $\simeq 6,000$ CPU hours. 
The initial fixed mesh described in Section 3.1 was used for approximately $1$s. After that, the  mesh
refinement was turned on.  The number of cells used in the simulation grew from $\simeq 5,000,000$ in the
beginning to 
$\simeq  30,000,000$  at the end of the simulation. 

Figure 1 presents the evolution with time of the  released nuclear energy $E_n$, kinetic
energy $E_k$,  binding energy  $E_{tot}$, and 
 burned mass $M_b$ inside the WD. The figure  shows that the deflagration  resulted in a  SNIa explosion. 
At $1.8$s, about $1.3\times 10^{51}$ ergs has already
been  released and  $E_{tot}\simeq 8\times 10^{50}$ergs is positive (the WD is unbound). 
At this moment,   the
WD radius is
$R_{WD}\simeq 4.7\times 10^8$ cm,  
$\rho_c \simeq 7\times 10^7$g/cc and the
expansion velocity 
$U_{exp} \simeq 1.0\times 10^9$ cm/s. 
Burning will probably release a few more units of
$10^{50}$ergs  until central density reaches
$\simeq  10^6$g/cc and burning quenches completely. 

Figure 2 shows the time evolution  of the flame surface. Initial 
 perturbations lead to  the formation of rising plumes of burned material. In the beginning, 
 the flame surface  and the energy release are small,
 and  the flame growth takes place  in essentially  hydrostatic conditions.
At $t<1$s,
the kinetic energy remains less that $\simeq 10^{47}$ ergs or $\simeq 0.01$\% of the binding
energy of the WD (Figure 1b).    The flame surface increases
significantly and becomes more wrinkled at later times. 
The  energy release rate increases, and the WD begins to expand. 

Figures 3 and 4 show the flame variable $f$  and the distribution of radial velocity 
$V_r$ at 
$1.44$s in three orthogonal planes, $x=3.2\times 10^7$cm, $y=3.2\times 10^7$cm, and $z=3.2\times 10^7$cm,
 offset  from the WD center.  
At this time the flame has approximately
reached the half-radius of the WD. The figures illustrate several important
features of burning. Outer parts of the flame consist of big plumes rising
through the WD. The absolute velocity of these plumes is $\simeq 6\times 10^8$cm/s or $\simeq a_s$. This
is about two times larger than the overall expansion velocity of matter,
$U_{exp}\simeq 3\times 10^8$cm/s. The  plume velocity relative to the
expanding material, $\simeq 0.5 a_s$, is subsonic.   Figure 4  also shows
regions  of unburned matter sinking to the center. There is a complex
flow pattern close to the WD center where fresh fuel is burned by the flame and  different parts of
the flame collide with each other and  annihilate. This  is similar to turbulent burning in a vertical
column in a uniform gravitational field (K95). In that case, the leading part of the flame is also
dominated by large scale bubbles whose motions determine the amount of fuel passing through the flame. The
fuel was burned inside the flame brush which is dominated by smaller-scale structures.  
Since the  scale of the
largest plumes in a column is  limited by the column width, the flame is able to reach a steady state on
all scales. In a spherical star, however, more space becomes available for the  plumes as the flame moves
to larger radii, and the rate of burning must increase with time.

Figure 5 shows  the effective turbulent flame speed $S_t$ which was estimated  as the speed of an
equivalent  spherical flame burning matter with the  same rate, 
$$
S_t \simeq {1\over 4\pi \rho r_f} \, {dM_b\over dt} ~,\eqno(15)
$$
where $r_f$ and $\rho_f$ are the average  radius and the density ahead of the  
flame
brush. In the simulation, a small-scale flame speed $S$ determined by (7) was 
$\simeq (1-2)\times 10^7$cm/s, depending on the radius, local density,
and time. Figures 1 and 5 show that burning rate  increased  at 
$t < 1.4-1.5$s and noticeably exceeded the value of $S$ due to the growth of the flame surface. 
However, $S_t$ reached maximum at $\simeq 1.5$s and started to decline at 
later times. The WD becomes unbound soon after maximum $S_t$ is reached
(see Fig.~1). 

Figures 6 and 7 show the flame and radial velocity at the end of
the simulation, $1.8$s. The leading part of the flame is still
dominated by large plumes, whereas  inner parts are wrinkled and contain  more complex structures. 
The outer edge of the flame is located at
$\simeq  0.8 R_{WD}$. The absolute velocity of the plumes has increased and reached $\simeq
10^9$cm/s. However, the average expansion velocity of the WD has  increased significantly and is
also
$U_{exp}
\simeq 10^9$cm/s. The velocity of the flame bubbles is essentially zero relative  to the
expanding matter, and  they have practically stopped rising. The expansion  completely froze
the RT-instability on large scales. Generation of  new flame surfaces on large scales 
ceased. Freezing through expansion   is the  reason for decrease  in
burning  rate at
$t>1.5$s.

There was an  earlier  three-dimensional computation of a deflagration explosion in a CO-WD 
(K95) that used a nonuniform expanding $185^3$ grid, assumed
 instantaneous transition of burned matter to the state of NSE, and  the same prescription for $S$
as in this paper.  This computation was terminated at
$\simeq 1.7$s. 
At
$\simeq 1.7$s, the simulation showed  plumes of  burned material rising at $\simeq 5\times
10^8$cm/s.  Approximately $35$\% of the matter was burned,
in agreement with this paper (Note that $M_b$ shown in Figure 16 of K95 must be
multiplied by $8$, as it  erroneously shows the amount of burned matter 
as per one octant of the star). In the earlier calculation, the WD is still slightly bound.

To illustrate mesh refinement,  Figure 8  shows the  mesh in the X-Y plane $z=3.2\times 10^7$cm at
$1.8$s. Figure 9 shows the mesh, density and velocity field in a small region in the X-Z plane
$y=3.2\times 10^7$cm with coordinates $x= 0-1.2\times 10^8$cm, $z=0-9.4\times 10^7$cm.  Despite  the
complexity of the flame surface and velocity field inside the flame brush, the overall expansion
of the WD  is remarkably one-dimensional. This is due  to the subsonic nature of burning which allows
pressure to  equilibrate inside the WD. Small deviations of the WD surface  from spherical symmetry are
only noticeable  at late times when the flame gets very  close to it. 

\bigskip\centerline{\it 3.3. Importance of small-scale burning}

\bigskip
Figure 10 shows the time evolution of $E_n$,  $E_k$,  $E_{tot}$, and 
 $M_b$ for the simulation of a $0.5C+0.5O$ WD with $S_{sub}=0$ in (6). 
All other
parameters were the same as in  the baseline case presented in  Section 3.2.  Flame
development was similar in the beginning. However, in this case, burning slowed down and then
practically stopped  because the expansion  decreased  $S_l$ by $\simeq 3$
 orders of magnitude.  There was no explosion in this
case and  the  WD remained bound. This shows that some amount of burning on unresolved scales must be
incorporated. Although $S\simeq 10^7$cm/s in the baseline
case was too small  to produce an explosion by itself, it is necessary to allow nuclear energy release
at a  flame front  whose area increases due to turbulence. A finite $S$  is also necessary to mimic 
the  reduction in flame surface  caused by  merging and burnout of 
small-scale structures. 
This is crucial in order to reproduce numerically a  self-similar regime of turbulent burning in which 
the burning rate  becomes independent of the details of flame behavior on small scales (K95). 
 Figure 11 compares the flame surface at $\simeq
1.3$s for both cases in the $X-Y$ plane passing through the center of the WD. This figure illustrates
the tendency of the flame to create more surface when $S$ is small and to create less surface when $S$
increases.  Simulations with more resolution and different prescriptions for $S$  
are required to determine  if we are indeed beginning to see the self-similar behavior of the flame.

\bigskip\centerline{\it 3.4. Nucleosynthesis}

\bigskip
Figure 12 shows  the distribution inside the WD  of Fe-group, 
Si-group (intermediate mass elements), and unburned CO matter at
$1.8$s. The location of Si-group elements is determined using the criterion
$\delta_{nse} < 0.95$ (equation (10)). The figure shows that unburned CO penetrates deep
inside the flame brush and is present in significant amounts   near the WD center. 
Incomplete burning
and the  formation of intermediate mass elements occur at densities less than  $\simeq 10^7$g/cc and
take place at radii greater than $\simeq 0.3 R_{WD}$. Closer to the center, the density is
still too high,  and burning produces of Fe-group nuclei. The production of
S-group nuclei will move inwards  as the WD expands further. Pockets of completely unburned CO are likely
to survive in central parts. Before burning quenches, one
 should also expect the formation of large amounts of  magnesium  in the central parts of the WD. In
one-dimensional models magnesium only forms immediately before quenching of incomplete burning in the
outer parts of the WD.  here, outermost parts of the WD are expected to stay completely unburned since the
penetration of the flame into the these layers has stopped (Section 3.2).

Figure 13 shows the distribution at $1.8$s of the neutron excess of matter $\eta=1=2Y_e$.
The maximum neutron excess, $\eta \simeq 0.013$, is much smaller than $\simeq 0.1$ typically obtained in 
one-dimensional SNIa simulations (Nomoto et al, 1984; Woosley \& Weaver, 1986; Khokhlov, 1991). This is
mostly due to a  reduction of the theoretical electron capture rates (Section 2.4). In addition, burned
matter spends less time near the center due to RT-mixing. Figure 13 shows that the high-$\eta$
products are present at radii up to 
$\simeq 0.8 R_{WD}$.

\bigskip\centerline{\it 3.5. Sensitivity to initial conditions} 

\bigskip
Figure 14 shows the results of a numerical experiment in which the location of the initial flame
bubble was offset from the WD center by ${1\over 3}$ of its radius ($\simeq 10^6$cm). The computational
domain was doubled, and a simplified flame propagation scheme
 with constant $q_f = q_{nse}-q_n(0)$ equal to the instantaneous NSE energy release was used.
Equations (8)-(11) were turned off to save  time.  A $30$\%  deviation of the ignition cite from
the center was enough to
cause a highly asymmetric flame. 

In another numerical experiment the initial carbon mass fraction $X_C$ was decreased from 
$X_C=0.5$ (Case B) to $X_C=0.2$. 
This resulted in less energy release and less buoyancy of burned products. Comparison between the
$X_C=0.5$ and $X_C=0.2$ runs (Figure 15) shows a significant  delay  and slower increase of burning.

\bigskip\centerline{\bf 4. Discussion and conclusions}

\bigskip
This paper described a physical model and an adaptive mesh refinement numerical algorithm for
three-dimensional modeling of Type Ia supernovae. It  also presented the first results of
three-dimensional   simulations using this model  of a deflagration in a nonrotating,
Chandrasekhar-mass,  carbon-oxygen white dwarfs.

The simulation of a $0.5C + 0.5O$ CO-WD with central density
$\rho=2\times 10^9$g/cc was  carried out for $\simeq 1.8$s. The leading edge of the flame was dominated
by large-scale, rising plumes of burned matter. Between these plumes, unburned CO was sinking towards the
WD center. Fine-structured flame surface developed  closer to the center below the plumes. The simulation
lasted long enough to  demonstrate  the effect of freezing  of large-scale turbulence due to expansion
predicted in (K95). The effective turbulent burning speed $S_t$  increased exponentially due to the
growth of the RT-instability,  reached maximum value $S_t\simeq 2\times 10^8$ cm/s at $\simeq 1.5$s,  and
then started to decline.   The expansion velocity of the WD reached
$\simeq 10,000$km/s by the end of the simulation, and  the motion of the buoyant plumes relative to
unburned matter practically stopped. 

The   deflagration resulted in a rather healthy
explosion.  About 60\% of the WD was burned by the end of the
simulation, and this released
$\simeq 1.3\times 10^{51}$ergs of nuclear energy. The total energy, $E_{tot}\simeq 8\times 10^{50}$ergs,
was positive and the WD was unbound.   By
$1.8$s, the WD radius was
$\simeq 4.7\times 10^8$cm, and burning still continued in its central parts.  It should be
expected that a few more units of
$10^{50}$ergs  would be released before burning would be quenched completely, so that the total kinetic
energy at infinity should  be in the range
$(1-1.3)\times 10^{51}$ergs. 
This is  $\simeq 50-80$\%  of the kinetic energy required to produce
a typical SNIa explosion. The total burned mass in this simulation is expected to be around $\simeq
1M\odot$. Detailed nucleosynthesis postprocessing is required to predict the amount of $^{56}Ni$
synthesized during the explosion. By analogy with one-dimensional simulations, we expect up to $\simeq 50$\%
of the burned mass, or $\simeq 0.5M_\odot$, to be 
$^{56}Ni$. This  is nearly enough to power a SNIa light curve. 

There are several 
 features of the computed explosion, however, that lead us to conclude that it is an incomplete
 model of a SNIa.  One distinct
feature  is the presence  of a massive,
$\simeq 0.4M_\odot$, outer layer of unburned CO surrounding  burned matter. The presence of such a layer
was  a typical feature of all one-dimensional deflagration models with  constant $S_t$ or with $S_t$
decreasing at the end of the explosion (Woosley \& Weaver, Khokhlov 1991). 
The exception is the W7 deflagration model (Nomoto et al., 1976) which created intermediate mass elements
at velocities up to
$\simeq 14,000$km/s by maintaining a high deflagration speed  and simultaneously
avoiding rapid expansion of the outer layers. The simulations presented here show the
decrease of the burning rate in the outer layers and are not consistent with the W7-like behavior.
The presence of the unburned outer layer
significantly limits the maximum  expansion velocities of the intermediate mass elements, 
 and this
presents a difficulty for the spectra of SNIa at maximum and before maximum light. 
The observations and spectral analyses indicate that
the line-forming region for the intermediate mass elements extends up to  
$\simeq 15,000$ and in some cases up to $\simeq 30,000$km/s in  velocity space  (Branch, 1981; Benetti
et al.  ,1991; Wells et al., 1994; H\"oflich,  1995; H\"oflich \& Khokhlov, 1996; Lentz et al. 2000).

The simulations also show that
the formation  of  intermediate mass elements in a three-dimensional deflagration explosion  occurs
throughout almost the entire WD,  and that some intermediate mass elements may form very close to the WD
center. This presents another difficulty for spectral modeling. The   spectra of normal bright SNIa
show
 that  the
 minimum velocity
 of the line-forming region of $Si$ is  $\simeq 8,000 - 11,000$km/s (H\"oflich \& Khokhlov, 1996),  $Ca$,
 $\simeq 4,000 - 6,000$km/s (Fisher et al., 1997), $Mg$, $\simeq 13,000-14,000$km/s (Meikle et al., 1996;
H\"oflich, 1997; Wheeler et al., 1998; Gerardy et al., 2000).  There is also an indication that the
minimum velocity of carbon in SN1990N (CII at 6580 A)  is larger than 
 26,000 km/s (Fisher et al. 1997), which means that practically all of the material in this supernova has
been burned. 
According to the simulations,  unburned CO is also likely to remain in the central regions of the 
deflagration supernova where it will coexist with $^{56}Ni$. Whether this can lead to the
excitation of
oxygen lines in the late-time spectra when the positron energy deposition becomes nonlocal (Milne et al.
1999)is an open question which requires further investigation (H\"oflich, private communication).

The formation of a massive outer layer of unburned material is a  large-scale effect caused by  the 
 competition between the of buoyancy of  large-scales structures (rising plumes) and  the
global expansion of the WD. In this competition,    buoyancy  eventually loses. 
During the explosion, 
the  expansion velocity of the star increases gradually,  and at $1.8$, outer layers have
velocities comparable  or even exceeding the velocity of the large-scale plumes
of burned products. Even if burning on small scales is underestimated in the simulation,  increasing the
energy released by small scales  would  increase the expansion velocity, but
would hardly affect the buoyancy of the plumes. As the WD continues to expand into the vacuum, the
rarefaction will  accelerate the outer layers  and this will further  increase their velocity relative to
the burned material.    

The difficulties associated with the  deflagration explosion can be avoided if burning turns into a
detonation during the explosion (delayed detonation). Detonation would easily overcome the expansion and
would incinerate the outer unburned layers of CO. This was one of the initial  reasons for the
introduction of the delayed detonation model of SNIa. It is also clear that the delayed detonation will
help with the potential  problems associated  with three-dimensional effects  by  wiping out 
 composition inhomogeneities in the inner layers. 

These are the first simulations with the model presented in the paper. As such, they show new effects but
also rais many questions. Computations with more resolution and with different subgrid burning models are
required to determine if the resolved scales  are small enough that burning on these scales is in  a
self-similar regime.  In particular, at maximum burning
($\simeq 1.5$s), the simulations show  large-scale motions with relative velocities of
$\simeq 5\times 10^8$cm/s. Thus there  should be shear instabilities and a strong turbulent cascade
on the sides of rising plumes. This should be incorporated into the model. The question of the
deflagration-to-detonation transition (DDT) must also be addressed.  A rough estimate can be made as
follows. Assume that at
$1.5$s, the turbulent velocity is $\simeq 10^8$cm/s on the driving scale $L\simeq 10^8$cm. Then, from 
Kolmogorov cascade, the turbulent velocity $U_\lambda \simeq 10^5\lambda$ cm/s on a scale $\lambda$, where
$\lambda$ is in centimeters. At $\rho\simeq 10^7$g/cc, the thickness of the laminar flame is
$\delta_l \simeq 10^2$cm, and on the scale $\lambda\simeq \delta_l$ the velocity $U_\lambda \simeq 10^7$cm
is much greater than $S_l$. Thus, the turbulent velocity on this scale may be strong enough to move the
flame from the flamelet into the distributed regime, create hot spots of nonuniformly preheated unburned
material,  and thus trigger a detonation (Khokhlov et al.,  1997). To trigger a detonation, the size of a
nonuniform region must exceed a critical value which depends on density. At $\rho\simeq 10^7$g/cc, the
critical size
$\simeq 10^6$cm is much less than $R_{WD}$ and the size of large scale structures obtained in the
simulations. Thus the transition to detonation seems to be possible. Significant work requires to
test these ideas. This paper shows that high-resolution three-dimensional
modeling of SNIa is now  a reality.

The calculations presented here indicate that  the result of the explosion may be  sensitive
to varying the position of the ignition site and the initial carbon fraction $X_C$ in a WD.
Decreasing of $X_C$ leads to less energy release and less buoyant products of burning. This results
in much 
slower development of the RT-instability, so that the freezing due to expansion  may be more
effective. On the other hand, smaller energy release leads to slower expansion as well.  
It is most likely that decreasing $X_C$
would result in a weaker explosion. Sensitivity to the position of the ignition site indicates that the
results may also be sensitive to stellar rotation,  which may lead to a preferential growth or suppression
of the initial perturbations along the rotational axis. This makes both $X_C$ and rotation potentially
good candidates for being the `hidden parameters' to describe the observed diversity of SNIa (Phillips,
1993).

\bigskip\noindent
{\it Acknowledgments.} I am grateful to Rob Rosenberg (NRL) and Almadena Chtchelkanova (Berkeley Research
Associates)  who developed
FTT-based tools for the surface rendering and visualization used in this work. This paper  benefited
greatly from discussions with Peter H\"oflich (UT Austin), Elaine Oran (NRL), Craig Wheeler (UT Austin),
Vadim Gamezo (Berkeley Research Associates), and Stirling Colgate (LANL). Work was supported in part by
the Naval Research Laboratory through the Office of Naval Research and by the NASA ATP program.

\vfill\eject
\bigskip\centerline{\bf References}

\hang\noindent
Arnett, W.D. \& Livne, E. 1994. ApJ 427, 314.

\hang\noindent
Benetti S., Cappellaro E., Turatto M. 1991, AA 247 410

\hang\noindent
Branch, D. 1981. ApJ, 248, 1076.

\hang\noindent
Colella, P. \& Glaz, H.M. 1985. J. Comput. Phys., 59, 264.

\hang\noindent
Colella, P. \& Woodward, P.R. 1984. J. Comput. Phys., 54, 174.

\hang\noindent
Fisher A., Branch D., H\"oflich P., Khokhlov A. 1995. ApJ 447L, 73.
 
\hang\noindent
Fisher A., Branch D., Nugent P., Baron E. 1997, ApJ 481L,  89.

\hang\noindent
Fowler, W.A., Caughlan, G.R. \& Zimmerman, B.A. 1975. ARA\&A, 13, 69. 

\hang\noindent
Gamezo, V.N.,   Khokhlov, A.M. \&  Oran, S.E.,
Proceedings of the 17th International Colloquium on the Dynamics of
Explosions and Reactive Systems, Universit\"at Heidelberg, IWR, 1999.

\hang\noindent
Gamezo, V.N.,  Vasil'ev, A.A.,  Khokhlov, A.M. \&  Oran, S.E.,
Proceedings of the 28th International Symposium on Combustion, vol.28, 2000 (in press).

\hang\noindent
Gerardy C., H\"oflich P., Fesen R., Wheeler J.C. 2000, AJ, to be submitted

\hang\noindent 
Hillebrandt, W. \& Niemeyer, J.C 2000. astro-ph/0006305 

\hang\noindent
H\"oflich, P. 1995. ApJ 443, 89.
 
\hang\noindent
H\"oflich P. 1997, in: Second Oak Ridge Symposium on  Atomic \& Nuclear
Astrophysics, ed. A. Mezzacappa, IOP Publishing, p. 693
 
\hang\noindent
H\"oflich \& Khokhlov 1996, ApJ 457, 500

\hang\noindent
Khokhlov, A.M. 1991. A\&A, 245, 114.

\hang\noindent
Khokhlov, A.M. 1995. ApJ, 449, 695.

\hang\noindent
Khokhlov, A.M., Oran, E.S., Wheeler, J.C. 1997. ApJ, 478, 678.

\hang\noindent
Khokhlov, A.M. 1998. J. Comput. Phys., 143, 519.

\hang\noindent
Khokhlov, A.M. \& Chtchelkanova, A.Yu. 1999. Proceedings of the 9-th SIAM Conference  on Parallel
Processing for Scientific Computing, San Antonio, TX, March 22-24, 1999. 

\hang\noindent
Khokhlov, A.M., H\"oflich, P.A., Oran, E.S., Wheeler, J.C., Wang, L. \& Chtchelkanova, A.Yu. 1999. ApJ,
524, L107.

\hang\noindent
Khokhlov, A.M. \& Oran, S.E. 1999. Combustion \& Flame, 119, 400.

\hang\noindent
Langanke, K. \& Martinez-Pinedo, G. 1999. Phys. Lett. B., 453, 187.

\hang\noindent
Lapidus, A. 1967. J. Comput. Phys. 2, 154.

\hang\noindent
Lentz, E.J., Baron, E., Branch, D., Hauschieldt, P.H. 2000. ApJ, sub\-mit\-ted; astro-ph/0007320.

\hang\noindent
Livne, E. 1993. ApJ 415, L17.

\hang\noindent
Martinez-Pinedo, G., Langanke, K. \& Dean, D.J. 2000. ApJ, 126:493. 

\hang\noindent
Meikle W.P.S., Cumming, , R.J., Geballe, T.R., et al. 1996. MNRAS 281, 263.

\hang\noindent
Milne P. A., The, L.-S.; Leising, M. D. 1999, ApJS 124, 503

\hang\noindent
Nomoto, K., Sugimoto, D. \& Neo, S. 1976. AP\&SS, 39, L37.

\hang\noindent
Nomoto, K., Thielemann, F.-K., Yokoi, K. 1984. ApJ, 286, 644.

\hang\noindent
Perlmutter, S., Aldering, G., Goldhaber, G., et al. 1999. ApJ, 517:565. 

\hang\noindent
Phillips, M.M. 1993. ApJ, 413, L105. 

\hang\noindent
Pskovskii, I. p. 1977. Soviet Astronomy, 21, 675.

\hang\noindent
Reinecke, M., Hillebrandt, W. \& Niemeyer, J.C. 2000. A\&A, in press.

\hang\noindent
Riess, A.G., Filippenko, A.V., Lui, M.C., et al. 2000. ApJ, 563:62.

\hang\noindent
Timmes, F. \& Woosley, S.E. 1992. ApJ, 396, 649.

\hang\noindent
Wells L.A., Phillips M.M., Suntzeff B. et al. 1994.  AJ 108, 2233.

\hang\noindent
Wheeler J.C., H\"oflich P., Harkness R., Spyromillo J. 1998,
   ApJ 496, 908 

\hang\noindent
Woosley, S.E. \& Weaver, T. 1986. ARA\&A, 24, 205.

\vfill\eject
\bigskip\centerline{\bf Figure captions}

\bigskip\item{Figure 1a. -}
Figure 1a. - Total ($E_{tot}$), kinetic ($E_k$), released nuclear ($E_n$)
energy and burned mass $M_b$ of a ${1\over 2}C+{1\over 2}O$ WD during a
deflagration explosion (Section 3.2).

\bigskip\item{Figure 1b. -}  
Solid line - kinetic energy ($E_k$) of a ${1\over 2}C+{1\over
2}O$ WD during a deflagration explosion at earlier times (Section 3.2).
Dashed line - kinetic energy with burning turned off (Section 3.1).

\bigskip\item{Figure 2a. -} Time evolution of the flame surface during a 
 deflagration explosion (Section 3.2).

\bigskip\item{Figure 2b. -} Same as  Fig. 2a. Details of the flame
structure at $1.8$s.

\bigskip\item{Figure 3. -} Flame variable $f$ (Equation (2)) for a
deflagration explosion Section 3.2 at
$1.44$s shown in three orthogonal planes offset by $3.2\times 10^7$cm
from the WD center. 

\bigskip\item{Figure 4. -} Same as Figure 3 but shows the radial velocity.

\bigskip\item{Figure 5. -} Estimated turbulent flame speed $S_t$
(Equation (15)) during a
deflagration explosion Section 3.2.

\bigskip\item{Figure 6. -} Flame variable $f$ (Equation (2)) for a
deflagration explosion Section 3.2 at the end of the simulation
$1.8$s shown in three orthogonal planes offset by $3.2\times 10^7$cm
from the WD center. 

\bigskip\item{Figure 7. -} Same as Figure 6 but shows the radial velocity.

\bigskip\item{Figure 8. -} Mesh and flame variable $f$ in the X-Y plane
passing through the WD center for a
deflagration explosion Section 3.2 at the end of the simulation
$t=1.8$s.

\bigskip\item{Figure 9a. -} Density and velocity in the X-Z plane
passing through the WD center for a
deflagration explosion Section 3.2 at the end of the simulation
$1.8$s. The region  $x=1.2\times 10^8$cm and $z=0-9.4\times 10^7$cm
is shown.

\bigskip\item{Figure 9b. -} Mesh in the X-Z plane
passing through the WD center for a
deflagration explosion Section 3.2 at the end of the simulation
$1.8$s. The region  $x=1.2\times 10^8$cm and $z=0-9.4\times 10^7$cm
is shown.

\bigskip\item{Figure 10. -} Time evolution of the 
total ($E_{tot}$), kinetic ($E_k$), released nuclear ($E_n$)
energy and burned mass $M_b$ of a ${1\over 2}C+{1\over 2}O$ WD during a
deflagration explosion with the laminar flame speed $S=S_l$ (Section 3.3).

\bigskip\item{Figure 11. -} Flame surface in the X-Y plane at $\simeq 1.3$s for the simulation with
(right panel, Section 3.2) and without (left panel, Section 3.3) turbulent small-scale burning.

\bigskip\item{Figure 12. -} Distribution of the Fe-group, Si-group, and CO matter for a
deflagration explosion Section 3.2 at the end of the simulation
$t=1.8$s shown in three orthogonal planes offset by $3.2\times 10^7$cm
from the WD center.

\bigskip\item{Figure 13. -} Same as Figure 12 but shows the neutron excess $\eta=1-2Y_e$.

\bigskip\item{Figure 14. -} Deflagration starting with the offset ignition point (Section 3.5).
Shows the flame variable and velocity at $\simeq 1$s. Black square shows the ignition point.

\bigskip\item{Figure 15. -} Evolution of the kinetic energy during deflagration explosion of a WD with
the initial
$0.5C + 0.5O$  and $0.2C+0.8O$ compositions (Section 3.5).

\end

1. what was done

2. large bubbles dominate flame growth, first rize fast then stop rizing -freesesot due to expansion
     probably cannot be bigger: large scales have no other mechanism but RT, small scales if increased,
most probably will add top expansion but not to plume rize. Examplifies major deflagration problem - how
to burn outer layers. This explosion is not W7.

3. final structure: unburned CO on top, mixture of everything below, deep penetration of CO to the
center, weird composition structure. Reduced neutronization.

\end